\documentclass[conference]{IEEEtran}
\IEEEoverridecommandlockouts

\usepackage{cite}
\usepackage{amsmath,amssymb,amsfonts}
\usepackage{algorithmic}
\usepackage{graphicx}
\usepackage{textcomp}
\usepackage{xcolor}
\usepackage{cutwin} 
\usepackage{caption} 
\usepackage[T1]{fontenc}

\usepackage[colorlinks=true, linkcolor=black, citecolor=black, urlcolor=black]{hyperref}

\usepackage[capitalise]{cleveref}
\usepackage{soul}

\def\BibTeX{{\rm B\kern-.05em{\sc i\kern-.025em b}\kern-.08em
    T\kern-.1667em\lower.7ex\hbox{E}\kern-.125emX}}

\begin{document}

\title{Control Protocol for Entangled Pair Verification in Quantum Optical Networks\vspace{-0.5em}}

\author{Vivek Vasan\IEEEauthorrefmark{2}, Anuj Agrawal\IEEEauthorrefmark{2}, Alexander Nico-Katz\IEEEauthorrefmark{3}$^{,}$\IEEEauthorrefmark{1}, Jerry Horgan\IEEEauthorrefmark{4}, \\Boulat A. Bash\IEEEauthorrefmark{5}, Daniel C. Kilper\IEEEauthorrefmark{4}, and Marco Ruffini\IEEEauthorrefmark{2}
	\vspace{0.5em}
	\\ \IEEEauthorrefmark{2}School of Computer Science and Statistics, CONNECT Centre, Trinity College Dublin, Dublin, Ireland\\\IEEEauthorrefmark{3}School of Physics, Trinity College Dublin, Dublin, Ireland\\\IEEEauthorrefmark{4}Department of Electronic and Electrical Engineering, CONNECT Centre, Trinity College Dublin, Dublin, Ireland\\\IEEEauthorrefmark{5}Department of Electrical and Computer Engineering, University of Arizona, Tucson, AZ, USA\\\IEEEauthorrefmark{1}Trinity Quantum Alliance, Unit 16, Trinity Technology and Enterprise Centre, Pearse Street, Dublin 2, Ireland \\
		\textit{Corresponding author email}: vasanv@tcd.ie
\vspace{-1em}}

\maketitle
               
\begin{abstract}
We consider quantum networks, where entangled-photon pairs are distributed using fibre optic links from a centralized source to entangling nodes.
The entanglement is then stored (via an entanglement swap) in entangling nodes' quantum memories until used in, e.g., distributed quantum computing, quantum key distribution, quantum sensing, and other applications.
Due to the fibre loss, some photons are lost in transmission.
Noise in the transmission link and the quantum memory also reduces fidelity. 
Thus, entangling nodes must keep updated records of photon-pair arrivals to each destination, and their use by the applications.
This coordination requires classical information exchange between each entangled node pair. 
However, the same fibre link may not admit both classical and quantum transmissions, as the classical channels can generate enough noise (i.e., via spontaneous Raman scattering) to make the quantum link unusable. Here, we consider coordinating entanglement distribution using a standard Internet protocol (IP) network instead, and propose a control protocol to enable such. We analyse the increase in latency from transmission over an IP network, together with the effect of photon loss, quantum memory noise and buffer size, to determine the fidelity and rate of entangled pairs. We characterize the relationship between the latency of the non-ideal IP network and the decoherence time of the quantum memories, providing a comparison of promising quantum memory technologies. 

\end{abstract}

\section{Introduction}
\label{sec:intro}


Quantum networks can revolutionize the way we transmit information securely and efficiently. By exploiting the principles of quantum mechanics, these networks enable applications such as quantum cryptography, distributed quantum computing, and high-precision sensing\cite{van2014quantum}. A fundamental resource in these networks is \emph{quantum entanglement}, when multiple particles become strongly correlated and the state of any subsystem cannot be regarded separately from the entire system. This leads to correlation between the particles that extends beyond those accounted by classical statistics. Entanglement can be distributed across distant nodes in a quantum network by optical photons, as they are highly resistant to environmental noise. These photons can be readily entangled with other photons, or atomic and solid state quantum memories, enabling entanglement consumption by various applications. 

The practical implementation of  quantum optical networks for entanglement distribution faces significant challenges due to losses and noise. For example, photons can be absorbed or scattered as they travel through optical fibers. Additionally, environmental noise can degrade the quality of the quantum bits (qubits). 
Unlike classical signals, quantum signals cannot be amplified because of the \emph{no-cloning theorem}, which prohibits creating identical copies of an unknown quantum state. This challenges the maintenance of entanglement over long distances or through complex network topologies.

Furthermore, when one photon of an entangled pair is lost in transmission, the remaining photon is unusable. This necessitates a protocol to manage entanglement distribution and storage. Given that quantum memories, which store qubits until needed, have limited capacity and finite coherence times, efficient resource management is crucial. The high arrival rate of photons relative to available memory necessitates that nodes promptly identify and discard unusable qubits in the memory to free up space for transferring entanglement from incoming photons into memory qubits. In addition, quantum applications require entangled pairs to maintain a certain fidelity between nodes. Therefore, each node must verify that its qubits have counterparts in the partner node’s quantum memory with the required fidelity before they can be served to application.

Here, we introduce a quantum optical network \emph{control protocol}. Existing protocols \cite{dahlberg2019link} are for entanglement identification and tracking in a meet-in-the-middle architecture, using a reactive entanglement distribution strategy  \cite{illiano2022quantum}.
Contrarily, we adopt a proactive  strategy using a midpoint-source architecture, where entangled photon pairs are continuously distributed to ensure an uninterrupted entanglement supply.
Our protocol enables nodes to verify the integrity of qubits and accurately pair them with their entangled counterparts. 
We analyse the network latency impact on the verification protocol, and, ultimately, on the entangled qubits' fidelity. This is important, as verification should ideally use the existing Internet protocol (IP) networks, rather than the quantum fibre link, as the  noise generated by the classical channel \cite{bahrani2018wavelength} could render the quantum channel unable to distribute entanglement. 



Indeed, latency impacts the execution time of the protocol and, as a result, the quality of entangled qubits. Quantum systems, including memories, are susceptible to decoherence, and parameters such as $T_1$ (longitudinal relaxation time, which measures the duration for spontaneous decay from the excited to the ground state) and $T_2$ (transverse relaxation time, which quantifies how long phase coherence is maintained in a superposition state) dictate how long qubits can preserve their coherence \cite{dasgupta2023adaptive}.  Classical control delays can thus degrade the fidelity of entangled pairs by increasing the verification time.

We analyze how this latency affects the fidelity of the entangled qubits, the rate of entanglement (i.e., the rate at which qubits with a maximally-entangled bipartite state are retained in the memory after verification), and the consumption of quantum memory resources needed to buffer entangled qubits until protocol completion. This analysis considers latency and noisy quantum memories under practical network conditions.

The remainder of the paper is organized as follows: \cref{sec:protocol and network model} provides an overview of the network model, integrating classical communication infrastructure with quantum networks.
\cref{sec:ent-degradation} describes our quantum memory noise model using Lindblad dynamics. \cref{sec:results} discusses the results of our study, and concludes the paper with a summary of our findings.

\section{Network Model and Control Protocol}
\label{sec:protocol and network model}
\subsection{IP Network for Classical Control Information Exchange}\label{AA}
\label{sec:IPnet}
Classical communication is essential for the operation of quantum networks, as it enables the exchange of measurement results required for quantum protocols, and control data needed for tasks like routing and resource management.

\begin{figure}[!ht]
    \centering
    \includegraphics[width=\linewidth]{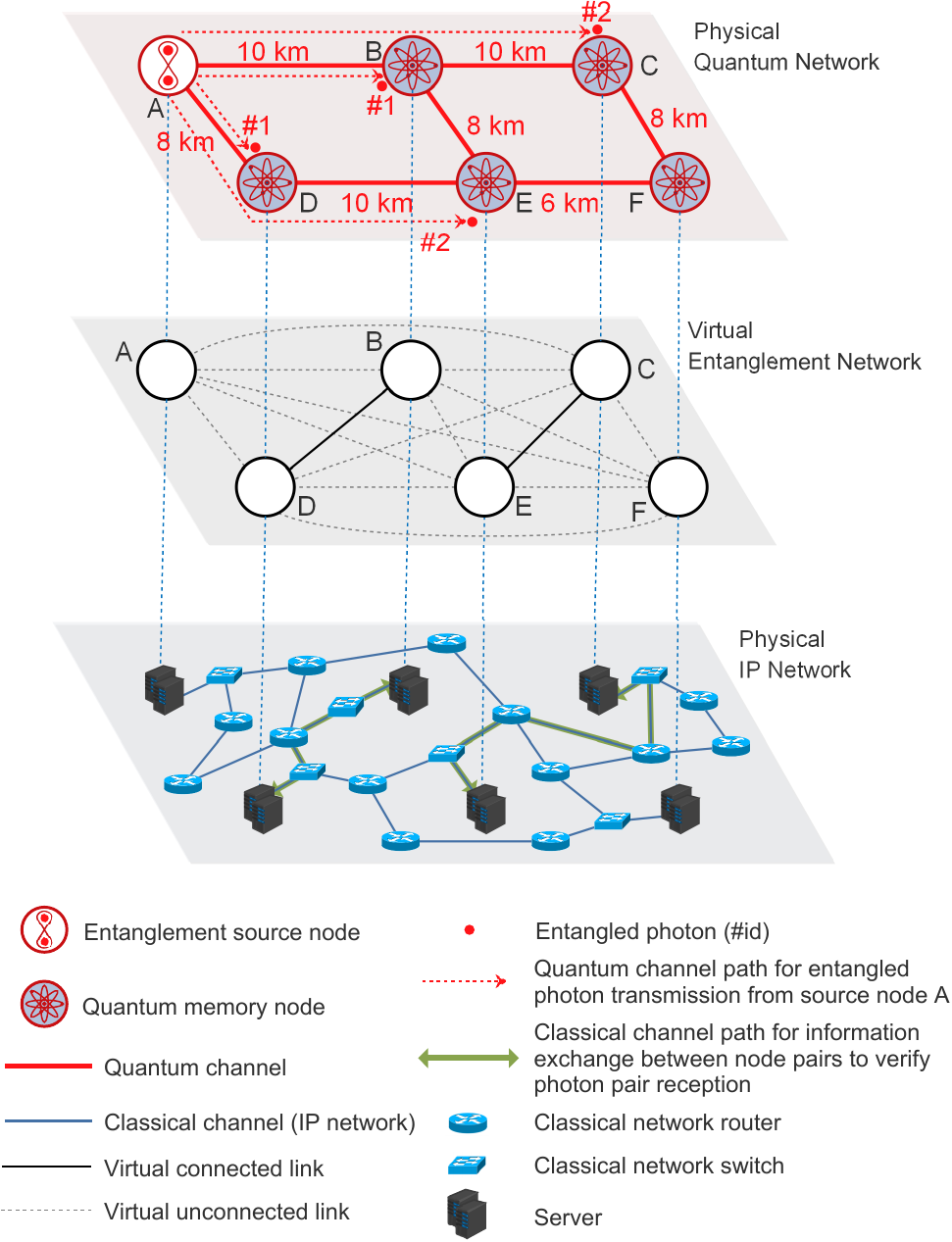}
    \caption{Quantum network integrated with IP network. After receiving entangled qubits, the entangling nodes exchange control information via IP network.}
    \label{fig:Network_fig}
    \vspace{-1.5em}
\end{figure}

Fig. \ref{fig:Network_fig} illustrates a quantum network with nodes having different capabilities such as entanglement generation, storage, routing, etc. Each node has two interfaces: one to a quantum transmission network (shown on the top layer of the figure), and one to an IP network (shown on the bottom layer of the figure). 
Internet traffic routes can take a different, often less direct path to the quantum routes, and congestion on the routers means that the latency will be variable. 

\subsection{Classical Headers with Heralded Source Information for Entangled Pair Identification} 
\label{sec:header}
A key requirement for entanglement distribution networks is to attach an identifier to entangled pairs at the source, so that they can be recognised and paired correctly at the destination. At minimum, this identifier or header should contain a sequence number that unequivocally identifies photons from that source. Additional information can include other parameters, such as exact photon wavelength (i.e., if required for mode conversion when mapping to a quantum memory \cite{chen2023zero, shapiro2024entanglement}), destination address (e.g., if no other means are available to identify the destination node), etc. In this work we assume such identification mechanism is already in place.  
For example, this could be a high-fidelity, low-loss scheme for entanglement distribution from \cite{chen2023zero, shapiro2024entanglement}.  It utilizes dual spectrally-multiplexed spontaneous parametric down-conversion (SPDC) sources to generate entangled biphotons. Detection of idler photons via Bell state measurements across dense wavelength-division multiplexing (DWDM) quantum channels heralds the corresponding signal photons. Classical information, including the frequency, and quantum state of the entangled signal known from Bell state measurement, is then shared with the entangling nodes along with the entangled photons by generating and transmitting header at a distinct (preferably longer) wavelength from the entangled photons. This  minimizes the risk of the header overlapping with entangled photons caused by chromatic dispersion. Additionally, the pump pulse repetition rate should be controlled to prevent interference. Each header is transmitted shortly before its corresponding entangled photon, maintaining a deliberate time gap between them. By using this strategy, each node can unambiguously link a received photon to its associated header and also measure the photon loss rate, as the classical headers are rarely lost.
The entangling nodes use this classical information on quantum state and frequency to mode-convert the photon to stationary qubit, while the arrival sequence determines the quantum memory slot used for storage. 
\subsection{Control Protocol}
\label{subsec:control protocol}
Here, we present a detailed control protocol designed to enable coordination between nodes for the identification and verification of entangled qubits in a quantum network. The protocol addresses challenges such as photon loss, finite quantum memory lifetime, and network-induced latency, ensuring efficient use of quantum resources while maintaining high fidelity of entanglement.

\subsubsection{\textbf{Protocol Overview}}

The proposed protocol operates alongside the entanglement distribution process described earlier. An entangled photon source generates pairs of entangled photons at specific intervals, that are directed towards quantum memories at the entangling nodes, as shown in \cref{fig:schematic}. Once transferred into the memory, each memory qubit is assigned a unique \emph{entanglement ID} based on the sequential ID number provided at the source, as detailed in \cref{sec:header}.

Due to transmission losses in optical fibers and optical routing elements, such as wavelength selective switches (WSSes), most photons are lost. Their arrival times may also vary slightly, due to environmental conditions affecting fibre propagation, imprecision in timing synchronisation across the network, and intrinsic quantum noise. %
The control protocol facilitates the exchange of entanglement IDs and status updates between the nodes, allowing them to match entangled pairs, manage their quantum memories efficiently, and maintain high-fidelity quantum entanglement. Only after this verification process is completed, can the entangled pairs be used by the application (i.e., distributed quantum computing, etc.). However, since the qubits in the memories decohere, timely operation of the verification protocol is important.

\subsubsection{\textbf{Protocol Steps}}
\label{sec:protocol-steps}
\underline{Photon detection and storage}:
On arrival, each photon interacts with a memory qubit, swapping its entanglement into the memory for storage at the respective node. The qubit classical header, traveling in the same fibre as the photon, is demultiplexed and decoded. The node records the associated entanglement ID. 
Each node maintains a local list of stored qubits and their entanglement IDs, with arrival timestamps  for timeout management. 

\underline{Entanglement ID exchange}:
When a node stores a new qubit, it sends a classical message containing the entanglement ID to the partner node via the IP network.  Both qubits in the entangled pair share the same identifier. 
This ensures that each entangled pair is distinctly identified and tracked throughout the communication process. We note that this occurs independently in the nodes seeking entanglement: each sends a message as soon as they store a qubit in their memory.

\underline{Matching and verification}:
When a node receives a message with an entanglement ID from its partner, it checks the existence of a qubit with the same entanglement ID in its memory. If the matching qubit is present, the node flags it in its memory as ready for quantum applications. The same steps are followed for the partner node and when both the nodes flag, reception of both the entangled qubits from the source is verified. 

\underline{Timeout and qubit discarding}:
A timeout $\Delta t$ for the protocol is established based on the quantum memory's transverse relaxation time $T_2$ (defined in \cref{sec:ent-degradation}) and the acceptable fidelity threshold $F_{th}$. 
The fidelity of stored qubits decays over time due to decoherence. We follow the modeling of fidelity decay over time in \cite{munro2015inside}, to determine the timeout period $\Delta t$ as 
%
%
        \begin{equation}
            \Delta t = -T_2 \cdot \ln\left( 2F_{th} - 1 \right),
            \label{eq:time_elapsed}
        \end{equation}
where the initial fidelity is assumed to be unity at the onset of decay. \cref{eq:time_elapsed} holds when $T_1 \gg T_2 \sim T_C$, where $T_C$ is the classical network latency between entangling node-pairs (see \cref{sec:ent-degradation}).

To maintain the fidelity of quantum information and efficiently manage the scarce memory resources, qubits must be discarded once their fidelity falls below the acceptable threshold, so that memory slots can be reused. This discarding process is streamlined by inferring entanglement IDs, capitalizing on the fact that entangled qubits in memory are stored in the same sequence that the entangled photons were generated at the source. Here is how the mechanism operates:
\begin{itemize}
\item Unmatched entanglement ID:
Since entanglement IDs are sequential, a node can infer which photons have not arrived based on the entanglement IDs of the subsequently received photons. Messages with entanglement IDs of the unreceived photons are exchanged between partner nodes and they discard the stored qubits (if any)  with those entanglement IDs. 
This ensures that only qubits with confirmed entangled partners are retained.

\item Timeout:  A node discards a qubit that exceed the set timeout period and sends a \emph{discard notification} containing the entanglement ID to its partner node, where a stored qubit with the same enatnglement ID is also discarded. 
\end{itemize} 

\section{Entanglement degradation due to idling in memory}

\label{sec:ent-degradation}

\begin{figure}[!ht]
    \centering
    \vspace{-0.5em}
    \includegraphics[width=0.99\linewidth]{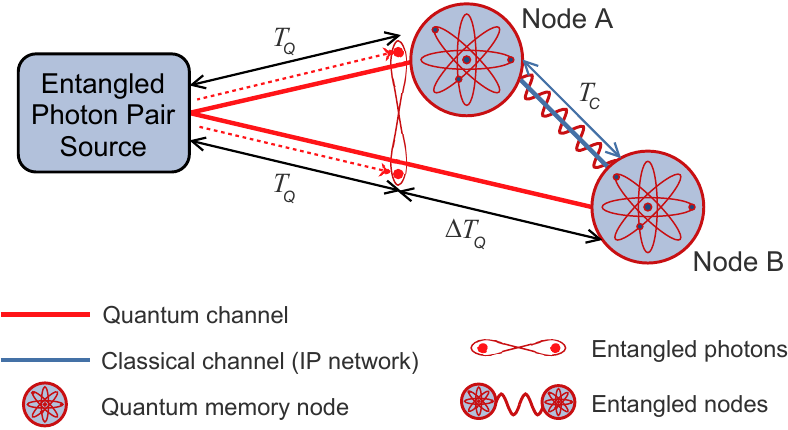}
    \caption{This depicts a scenario where an entangled photon pair is distributed to node A and node B. Due to the difference in channel lengths from source to node A and B, when a photon arrives at node A after time $T_Q$, its paired photon is still in flight and it reaches node B after time $T_Q+\Delta T_{Q}$. $T_C$ is the latency of classical channel (IP network) over which information exchange happens to verify photon pair reception.}
    \label{fig:schematic}
    \vspace{-1.7em}
\end{figure}
\subsection{Noise and Quantum State Fidelity}
\label{sec:fidelity}
We consider, as discussed in \cref{subsec:control protocol} and shown schematically in \cref{fig:schematic}, the framework, wherein each photon in an entangled pair (here, the Bell singlet) is transmitted via optical fibre to two nodes, A and B, respectively. The time taken $\Delta T$ for the protocol to complete verification for each entangled pair of qubits is given by $\Delta T = \Delta T_Q + T_C$,
where $\Delta T_Q$ represents the time discrepancy caused by unequal quantum channel lengths, resulting in one qubit being exposed to noise for a longer duration than its counterpart, and $T_C$ is the classical verification protocol runtime.
We make two additional reasonable assumptions, which focus our analysis. 

First, signaling times in the classical network are symmetric $T_{A\to B} = T_{B \to A} = T_C$. This is justfied by the bidirectional fibre links consisting of fibre pairs. We note that, if considering individual packets, latency can be different across the two directions due to local queuing at the routers. However, in this case, we can still  consider the larger of the two latencies. Second, we assume that fiber noise has negligible impact on the fidelity of the entangled qubits, for the following reasons. The effect of quantum noise in optical fibers depends on the optical degree of freedom used to encode entanglement. While polarization entanglement is susceptible to polarization mode dispersion caused by birefringence in optical fibers, recent study \cite{craddock2024automated} has demonstrated that this can be mitigated using polarization compensation systems, enabling high-fidelity entanglement distribution. Alternatively, converting polarization into time-bin entanglement before transmission is effective \cite{shapiro2024entanglement}, as time-bin encoding is robust over long-distance transmission \cite{kim2024fully}. Thus, we focus solely on the the quantum memory noise. Our figure of merit, which characterizes the degradation of entanglement over time, is the fidelity of $\hat{\rho}(t)$ with the initial, maximally-entangled, Bell singlet $\hat{\rho}(0) = |\Psi^-\rangle\langle \Psi^-|$\cite{nielsen2010quantum}:
\begin{equation}\label{eq:fidelity}
    \mathcal{F}(t) = \langle\Psi^-|\hat{\rho}(t)|\Psi^-\rangle.
\end{equation}
\subsection{Constructing the Lindblad Master Equation}
\label{sec:lindblad}

Quantum memory registers exhibit amplitude damping and phase errors governed by their respective $T_1$ and $T_2$ times, 
which we assume to be identical at all nodes, although an extension to node-dependent decoherence times is straightforward. 
We now construct a Lindblad equation which accounts for the degradation of the quantum state over time, subjected to these errors, whilst it idles in memory.

Overall, the quantum channel acting on memories in nodes A and B consists of the unitary evolution under the Hamiltonian $\hat{H} = \mathbb{I}_A\otimes\mathbb{I}_B$, corresponding to an ideal memory realizing trivial evolution, i.e., idling. This ideal behaviour is interrupted by stochastic amplitude damping and phase errors governed by the jump operators:
\begin{equation}
    \hat{L}_{1,A(B)} = \sqrt{\gamma_1} \hat{\sigma}^-_{A(B)}, \quad \hat{L}_{2,A(B)} = \sqrt{\gamma_2} \hat{\sigma}^z_{A(B)}
\end{equation}
where the decay rates are given by $\gamma_{j} = 1/T_j$, $j\in\{1,2\}$ \cite{singh2020using}, and  $\hat{\sigma}^z_{A(B)}$ and $\hat{\sigma}^-_{A(B)}$ are the standard Pauli $z$-operator and lowering operators, acting on node A(B), respectively.

The Lindblad master equation governs the time evolution of the (generically mixed) quantum state $\hat{\rho}(t)$ and is \cite{breuer2002theory}:
\begin{equation}\label{eqn:lindblad}
    \partial_t \hat{\rho}(t) = -i[\hat{H}, \hat{\rho}(t)] + \sum D(\hat{L}_{j,\alpha}, \hat{\rho}(t)),
\end{equation}
where $i=\sqrt{-1}$, the sum is over $j\in\{1,2\}$ and $\alpha\in\{A, B\}$, the commutator is $[\hat{H}, \hat{\rho}(t)] = \hat{H}\hat{\rho}(t)-\hat{\rho}(t)\hat{H}$, and the dissipators $D(\hat{L}_{j,\alpha}, \hat{\rho}(t)) = \hat{L}_{j,\alpha} \hat{\rho}(t) \hat{L}_{j,\alpha}^\dagger - \frac{1}{2} \{ \hat{L}_{j,\alpha}^\dagger \hat{L}_{j,\alpha}, \hat{\rho}(t) \}$ account for the effect of the jump operators. The anti-commutator is  $\{ \hat{L}_{j,\alpha}^\dagger \hat{L}_{j,\alpha}, \hat{\rho}(t) \} = \hat{L}_{j,\alpha}^\dagger \hat{L}_{j,\alpha} \hat{\rho}(t) + \hat{\rho}(t) \hat{L}_{j,\alpha}^\dagger \hat{L}_{j,\alpha}$. 
We assume that the initial state the Bell singlet $\hat{\rho}(0)$. In practice, it may be, e.g., a Bell-diagonal mixed state \cite{jones2020exploring} having initial fidelity with $\hat{\rho}(0)$ informed by decoherence in the fiber. As we focus on degradation during idling in memory, we defer a complete analysis of these to the future.

The average master equation form of \cref{eqn:lindblad} describes a system experiencing Markovian dynamics mediated by continuous (weak) measurement under the jump operators. In the context of our work, wherein $T_1$ and $T_2$ times are large relative to protocol latency, the most likely situation is that neither error occurs during idling.
It is instructive to consider an alternative approach to the full density-matrix propagation of \cref{eqn:lindblad}: the `unraveling' of \cref{eqn:lindblad} into individual trajectories. This unraveling procedure involves pure state propagation of the initial state, subject to random applications of the $\hat{\sigma}^z_{A(B)}$ and $\hat{\sigma}^-_{A(B)}$ at times dictated by the corresponding error rates $\gamma_1$ and $\gamma_2$ respectively. The resulting states $|\psi_r(t)\rangle$, when averaged over many runs $N_\text{traj}$ such that $\hat{\rho}(t) = \sum_r | \psi_r(t)\rangle\langle \psi_r|/N_\text{traj}$, yield identical results to full integration of the Lindblad master equation in the limit $N_\text{traj} \to \infty$. Technical details are in \cite{barchielli2009quantum}. In the specific context of this article: (i) errors are typically rare, and (ii) an error induces a change in the state which renders it orthogonal to the Bell singlet. Whilst we do not carry out this trajectory unraveling explicitly in this work, the above remarks bear relevance to the interpretation of the quantum state fidelity in \cref{sec:fidelity}: in the context of the above setup wherein jumps are relatively rare, \cref{eq:fidelity} corresponds identically to the probability of retaining, on any given trajectory, the perfect Bell state $|\Psi^-\rangle$ at time $t$.
\section{Results}
\label{sec:results}

Here, we numerically evaluate the performance of the proposed control protocol with respect to entangled pair rate and entanglement fidelity. We carry out multiple simulations, using different values for latency and different quantum memory relaxation times \( T_1 \) and \( T_2 \), corresponding to specific quantum memory technologies shown in \cref{tab:tab1}. Additionally, we examine the buffer size necessary for storing idling qubits during the protocol execution. Simulations are performed using NetSquid \cite{netsquid} except for the Lindblad dynamics which are implemented using QuTiP package \cite{qutip}. In order to decouple the protocol from the application, we also assume that entangled qubits are consumed as soon as the protocol verifies their integrity (and, in that case, their memory location is also reallocated).

\begin{table}[htbp]
\caption{Experimental parameters for different quantum memory technologies\vspace{-1em}}
\label{tab:tab1}
\begin{center}
\begin{tabular}{|c|c|c|}
\hline
\textbf{Quantum Memory Technology} & \multicolumn{2}{|c|}{\textbf{Parameters}} \\
\cline{2-3}
& \textbf{\textit{T\textsubscript{1} (s)}} & \textbf{\textit{T\textsubscript{2} (s)}} \\
\hline
Ion Trap (\(^{171}\text{Yb}^+\)) & 12000\textsuperscript{b} & 4200\textsuperscript{b} \\
\hline
Rare Earth Ions (\(^{167}\text{Er}^{3+}:Y_2SiO_5\)) & 600\textsuperscript{c} & 1.3\textsuperscript{c} \\
\hline
Ion Trap (\(^{40}\text{Ca}^+\)) & 1.14\textsuperscript{a} & 0.5\textsuperscript{a} \\
\hline
NV Centers in Diamond (Nuclear Spin) & 200\textsuperscript{d} & 0.5\textsuperscript{d} \\
\hline
Superconductor Cavity & 0.0256\textsuperscript{e} & 0.034\textsuperscript{e} \\
\hline
Superconductor Cavity & 0.0012\textsuperscript{f} & 0.00072\textsuperscript{f} \\
\hline
\multicolumn{3}{l}{\textsuperscript{a}\cite{kreuter2005experimental}, \textsuperscript{b}\cite{wang2021single}, \textsuperscript{c}\cite{rancic2018coherence}, \textsuperscript{d}\cite{maurer2012room}, \textsuperscript{e}\cite{milul2023superconducting}, \textsuperscript{f}\cite{reagor2016quantum}}
\end{tabular}
\end{center}
\vspace{-2em}
\end{table}

\begin{figure}[!ht]
    \centering
    \vspace{-0.5em}
    \includegraphics[width=\columnwidth]{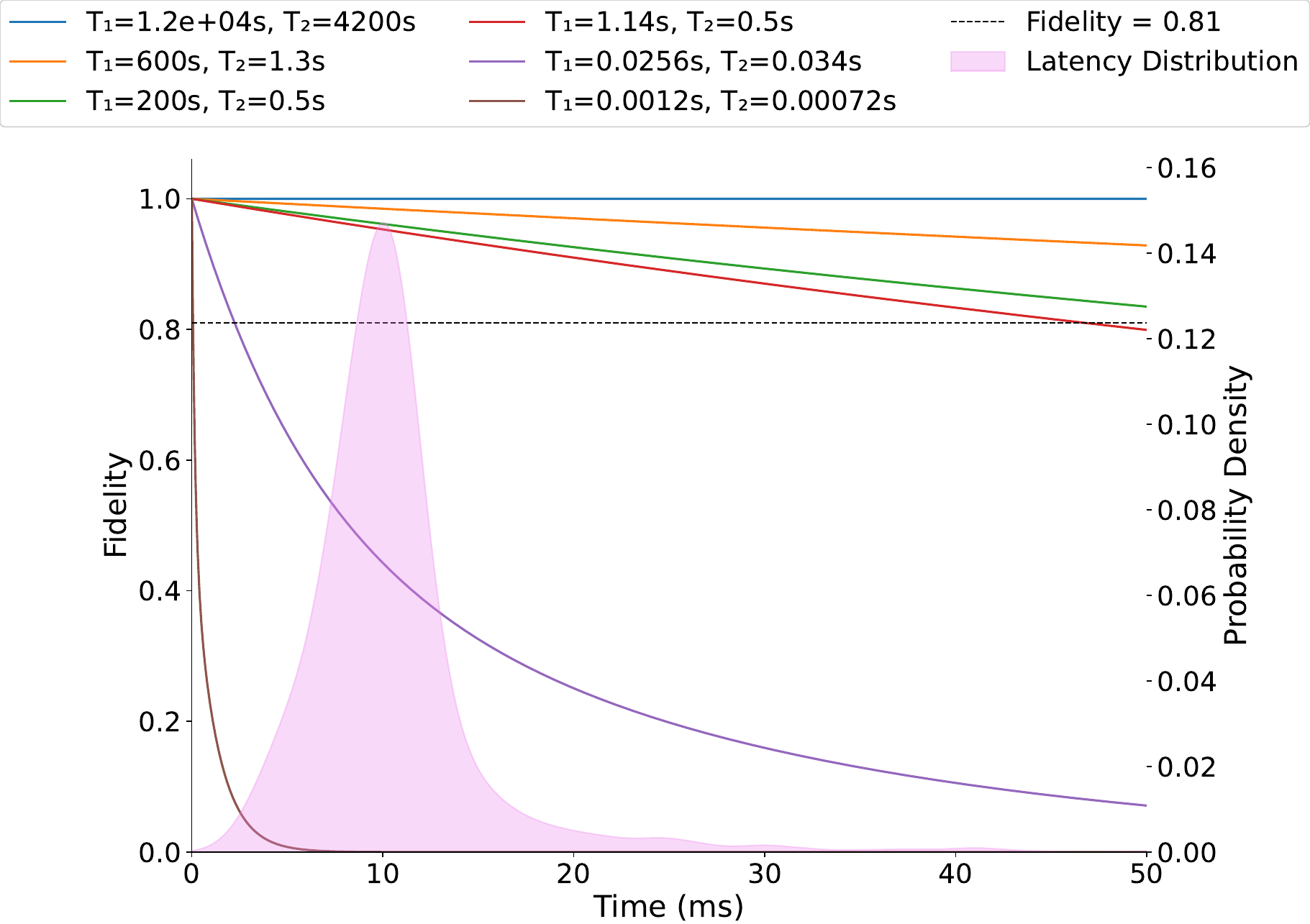}
    \vspace{-1.5em}
    \caption{\textbf{Degradation of state fidelity in realistic quantum memories}. Evolution of the quantum state fidelity $\mathcal{F}$ (left ordinate) with the Bell singlet state as a function of total classical latency time $T_C$. Different quantum memory technologies are characterized by $T_1$ and $T_2$ relaxation times sourced from \cref{tab:tab1}. Technologies are ordered by decreasing $T_2$ times, highlighting the impact of $T_2$-error dominance ($T_1 \gg T_2$).
    Dotted black line indicates the threshold value $\mathcal{F} = 0.81$ required for QKD. Shaded pink region shows the distribution of IP network latencies for the Dublin Metro Area. Probability density (right ordinate) is fitted to the empirical data extracted from \cite{ookla2024speedtest}.}
    \label{Fig:fidelity-fixed_latency}
    \vspace{-0.5em}
\end{figure}

The network topology used in the simulations is depicted in \cref{fig:Network_fig}. For the quantum network, as per the node architecture presented in \cite{bali2024routing}, the loss due to an intermediate node is considered to be 8 dB and the loss due to the source node and the memory node is considered to be 4 dB. Fibre attenuation is considered to be 0.2 dB/km.
To incorporate practical values for IP network latency, we utilize empirical data extracted for Q1 2024 for broadband network for Greater Dublin Area from Ookla's open data initiative \cite{ookla2024speedtest}. 
This dataset captures typical transmission delays in urban fiber networks, enabling our simulations to closely approximate practical metro-scale classical network performance.
In \cref{Fig:fidelity-fixed_latency}, we overlap the plot of the probability density function fitted to empirical distribution of network latency data (pink region) to the plot reporting the quantum state fidelity.

Fig. \ref{Fig:fidelity-fixed_latency} illustrates the fidelity decay for various $T_1$ and $T_2$ values based on the current experimental parameters of quantum memory technologies presented in \cref{tab:tab1}. Our analysis shows that state fidelity is more vulnerable to dephasing noise, governed by $T_2$, than to amplitude damping noise, which depends on $T_1$. However, when $T_2$ times are comparable across different systems, the influence of amplitude damping noise related to $T_1$ becomes significant.
Consider \(^{40}\text{Ca}^+\) ion trap memories with $T_1 = 1.14$~s and $T_2 = 0.5$~s, compared to an NV center in diamond with $T_1 = 200$~s and $T_2 = 0.5$~s. Despite having identical $T_2$ values, the ion trap experiences a faster fidelity decay due to its shorter $T_1$. This demonstrates that while $T_2$ times are typically shorter, and, thus, have a greater impact on fidelity, the effects of $T_1$ cannot be ignored. While the minimum fidelity requirements are application dependent, in order to highlight a realistic use case for today's level of development of networked quantum systems, we show a fidelity threshold of 81\% in \cref{Fig:fidelity-fixed_latency}, which is the minimum requirement for quantum key distribution (QKD)\cite{wengerowsky2018entanglement}. The latency distribution (pink region) for the Dublin-Metro area suggests that sufficiently high-fidelity can be maintained over typical IP network latencies, except for superconductor cavities. 


To quantitatively assess the demands on quantum memory resources, we analyze the required buffer sizes for storing idle qubits under varying network conditions. 
\begin{figure}[!ht]
    \centering
    \vspace{-0.5em}
    \includegraphics[width=\columnwidth]{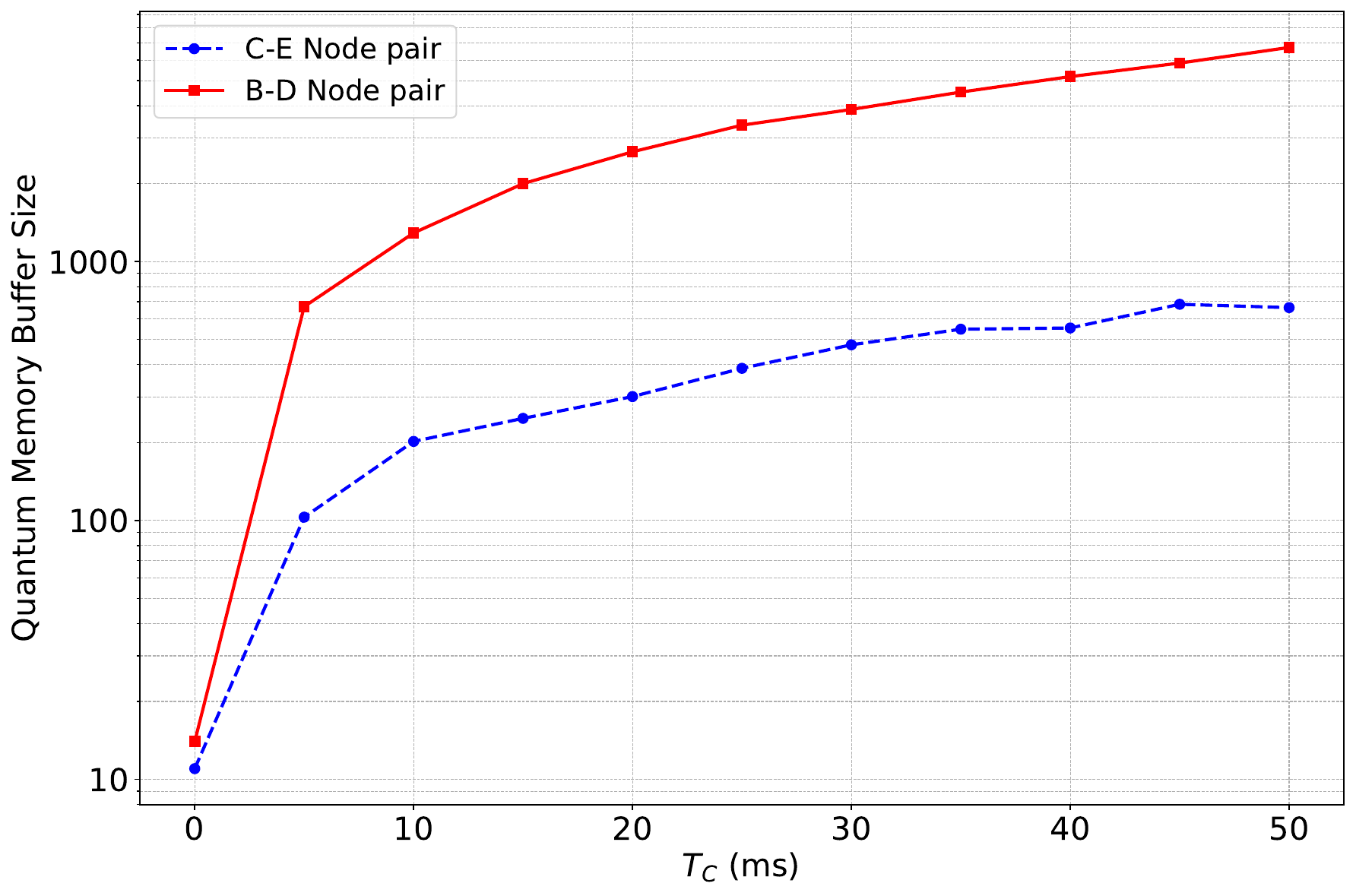}
    \vspace{-1.5em}
    \caption{\textbf{Buffer size required for idling qubits for different node pairs.} Increased network latency leads to higher consumption of quantum memory slots due to longer queuing delays of entangled qubits in the memory for $T_1 = 1.14~\text{s},~T_2 = 0.5~\text{s}$.}
    \label{fig:Buffer_size}
    \vspace{-1.5em}
\end{figure}

\begin{figure}[!ht]
    \centering
    \includegraphics[width=\columnwidth]{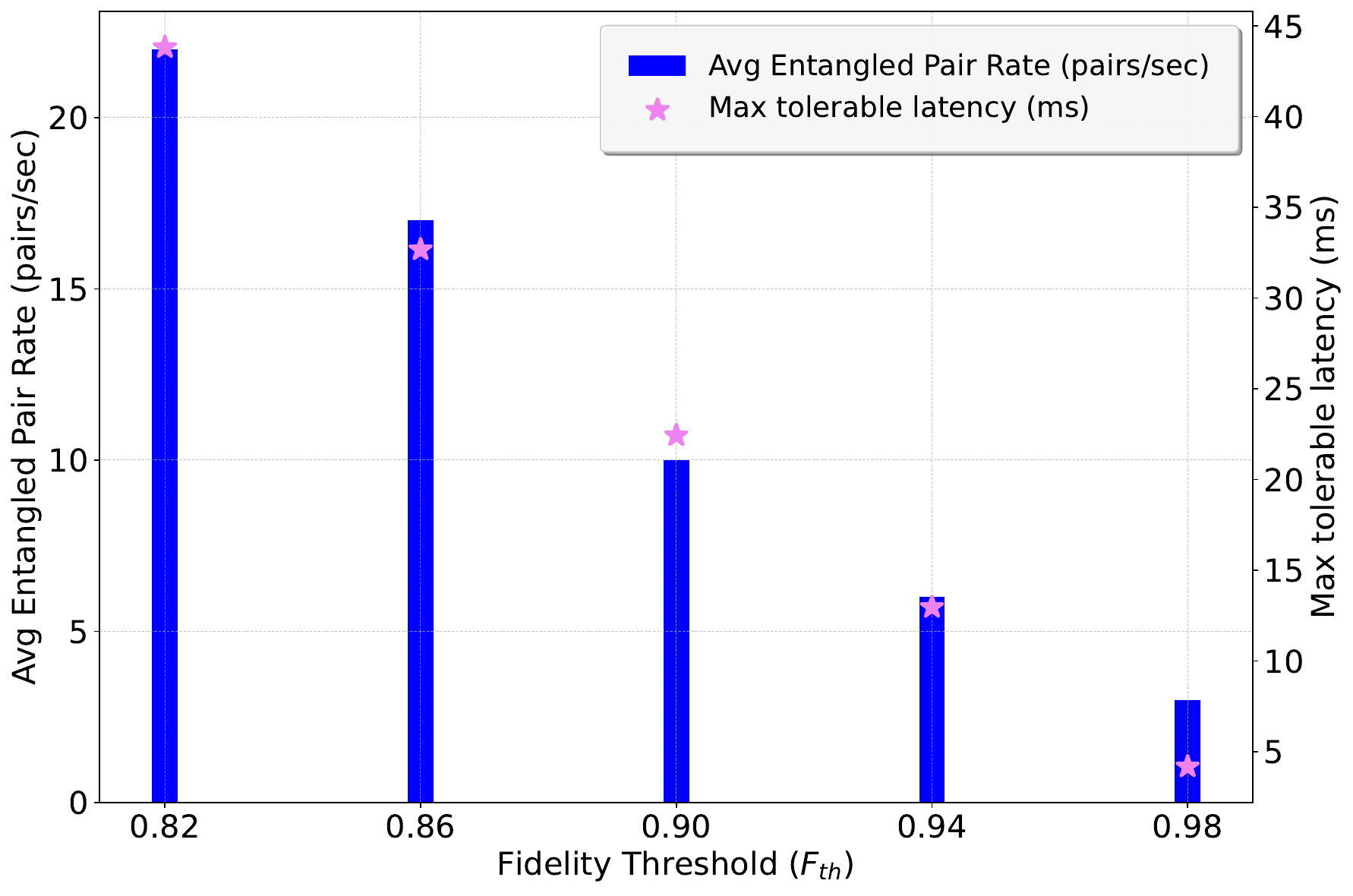}
    \vspace{-1.5em}
    \caption{\textbf{Impact of timeout duration on entangled pair rates across different fidelity thresholds}. The timeout for each fidelity threshold is determined by \cref{eq:time_elapsed} and maximum tolerable latency that maintains fidelity above the threshold is interpolated from the fidelity curve in \cref{Fig:fidelity-fixed_latency} for $T_1$ = 1.14 s and $T_2$ = 0.5 s. The result is shown for C-E node pair in \cref{fig:Network_fig}.}
    \label{Fig:epr_rate}
    \vspace{-0.5em}
\end{figure}
In \cref{fig:Buffer_size}, we plot the buffer size required for our protocol to complete the verification process, providing concrete numerical data on buffer requirements under different conditions. The entangled photon generation rate at the source is set to 1.3\,MHz\cite{sansa2022visible}. Buffer occupation is influenced by three factors: the entangled pair generation rate at the source, the loss within the network, and the execution time of the protocol. For the specified destination node pairs (C-E and B-D in \cref{fig:Network_fig}) both the entanglement generation rate and time difference $\Delta T_Q$ remain constant, as they are determined by fixed parameters such as the quantum channel length and the characteristics of the source used. Consequently, the buffer size is only affected by network latency $\Delta T_C$, which directly impacts the protocol's completion time, and the loss that is dependent on number of intermediate nodes. \cref{fig:Buffer_size} indicates that increased latency necessitates larger buffer capacities to accommodate the qubits queuing and awaiting verification, as expected.


Finally, in \cref{Fig:epr_rate} we analyse the impact of timeout on the entangled pair rate across different values of fidelity threshold $\mathcal{F}_{th}$ at the entangling nodes. Higher $\mathcal{F}_{th}$ requires shorter timeouts, as the decoherence to the higher $\mathcal{F}_{th}$ happens faster, leading the protocol to discard qubits more frequently, which, in turn, reduces the rate. Thus, we obtain the maximum tolerable latency $T_{C,\max}$ such that fidelity threshold $\mathcal{F}(T_{C,\max})\geq\mathcal{F}_{th}$. \cref{Fig:epr_rate} highlights the trade-off between maintaining high fidelity and sustaining a high entangled-pair rate. Quantum memories with longer lifetimes maintain higher entangled pair rates without compromising the fidelity.

\section{Conclusion}
\label{sec:conclusion}
In this paper, we present a control protocol for coordinating nodes in a quantum network to verify and manage entangled qubits. By addressing photon loss, finite quantum memory lifetimes, and network-induced latency, our protocol optimizes the use of quantum resources while maintaining high entanglement fidelity.

Our study reveals that, beyond a certain threshold of quantum memory coherence time, the deployment of internet-based control signals becomes feasible without substantially compromising entanglement fidelity or entangled pair rates. Specifically, when memory coherence times surpass this critical value, the additional latency introduced by IP network-based control signals does not significantly degrade protocol performance. We have demonstrated that several quantum memory technologies listed in \cref{tab:tab1} meet the minimum requirements for our protocol to achieve the fidelity thresholds necessary for applications such as quantum key distribution (QKD) under typical metropolitan network latencies (10 ms). Moreover, in scenarios where latency occasionally extends to 20–30 ms, quantum memories with longer $T_2$ times become essential to maintain high-fidelity qubits during idling periods.

These findings have significant implications for the practical deployment of quantum networks. They affirm the viability of integrating quantum networks with existing IP infrastructures in metropolitan areas, provided that quantum memories meet or exceed the relevant coherence times. Achieving these memory lifetimes is crucial for ensuring that quantum networks can operate effectively over classical IP infrastructure, thus supporting the scalable deployment of quantum technologies within metro-scale networks.


\section*{Acknowledgment}
This material is based upon work supported by the Science Foundation Ireland grants 20/US/3708, 21/US-C2C/3750, and 13/RC/2077 P2 and National Science Foundation under Grant No. CNS-2107265.

\bibliographystyle{IEEEtran}
\bibliography{references}

\begin{thebibliography}{10}
\providecommand{\url}[1]{#1}
\csname url@samestyle\endcsname
\providecommand{\newblock}{\relax}
\providecommand{\bibinfo}[2]{#2}
\providecommand{\BIBentrySTDinterwordspacing}{\spaceskip=0pt\relax}
\providecommand{\BIBentryALTinterwordstretchfactor}{4}
\providecommand{\BIBentryALTinterwordspacing}{\spaceskip=\fontdimen2\font plus
\BIBentryALTinterwordstretchfactor\fontdimen3\font minus \fontdimen4\font\relax}
\providecommand{\BIBforeignlanguage}[2]{{%
\expandafter\ifx\csname l@#1\endcsname\relax
\typeout{** WARNING: IEEEtran.bst: No hyphenation pattern has been}%
\typeout{** loaded for the language `#1'. Using the pattern for}%
\typeout{** the default language instead.}%
\else
\language=\csname l@#1\endcsname
\fi
#2}}
\providecommand{\BIBdecl}{\relax}
\BIBdecl

\bibitem{van2014quantum}
R.~Van~Meter, \emph{Quantum Networking}.\hskip 1em plus 0.5em minus 0.4em\relax Hoboken, NJ, USA: Wiley, 2014.

\bibitem{dahlberg2019link}
A.~Dahlberg, M.~Skrzypczyk, T.~Coopmans, L.~Wubben, F.~Rozpędek \emph{et~al.}, ``A link layer protocol for quantum networks,'' in \emph{Proc. ACM SIGCOMM}, 2019, pp. 159--173.

\bibitem{illiano2022quantum}
J.~Illiano, M.~Caleffi, A.~Manzalini, and A.~S. Cacciapuoti, ``Quantum internet protocol stack: A comprehensive survey,'' \emph{Comput. Netw.}, vol. 213, p. 109092, 2022.

\bibitem{bahrani2018wavelength}
S.~Bahrani, M.~Razavi, and J.~A. Salehi, ``Wavelength assignment in hybrid quantum-classical networks,'' \emph{Sci. Rep.}, vol.~8, no.~1, p. 3456, 2018.

\bibitem{dasgupta2023adaptive}
S.~Dasgupta, T.~S. Humble, and A.~Danageozian, ``Adaptive mitigation of time-varying quantum noise,'' in \emph{Proc. IEEE Int. Conf. Quantum Comput. Eng. (QCE)}, 2023, pp. 99--110.

\bibitem{chen2023zero}
K.~C. Chen, P.~Dhara, M.~Heuck, Y.~Lee, W.~Dai, S.~Guha, and D.~Englund, ``Zero-added-loss entangled-photon multiplexing for ground- and space-based quantum networks,'' \emph{Phys. Rev. Appl.}, vol.~19, no.~5, p. 054029, 2023.

\bibitem{shapiro2024entanglement}
J.~H. Shapiro, M.~G. Raymer, C.~Embleton, F.~N.~C. Wong, and B.~J. Smith, ``Entanglement source and quantum memory analysis for zero-added-loss multiplexing,'' \emph{Phys. Rev. Appl.}, vol.~22, no.~4, p. 044014, 2024.

\bibitem{munro2015inside}
W.~J. Munro, K.~Azuma, K.~Tamaki, and K.~Nemoto, ``Inside quantum repeaters,'' \emph{IEEE J. Sel. Top. Quantum Electron.}, vol.~21, no.~3, pp. 78--90, 2015.

\bibitem{craddock2024automated}
A.~N. Craddock, A.~Lazenby, G.~B. Portmann, R.~Sekelsky, M.~Flament, and M.~Namazi, ``Automated distribution of polarization-entangled photons using deployed new york city fibers,'' \emph{PRX Quantum}, vol.~5, no.~3, p. 030330, 2024.

\bibitem{kim2024fully}
J.~Kim, J.~Park, H.-S. Kim, G.~Kim, J.~T. Kim \emph{et~al.}, ``Fully controllable time-bin entangled states distributed over 100-km single-mode fibers,'' \emph{EPJ Quantum Technol.}, vol.~11, no.~1, p.~53, 2024.

\bibitem{nielsen2010quantum}
M.~A. Nielsen and I.~L. Chuang, \emph{Quantum Computation and Quantum Information}.\hskip 1em plus 0.5em minus 0.4em\relax Cambridge, U.K.: Cambridge Univ. Press, 2010.

\bibitem{singh2020using}
H.~Singh, Arvind, and K.~Dorai, ``Using a lindbladian approach to model decoherence in two coupled nuclear spins via correlated phase damping and amplitude damping noise channels,'' \emph{Pramana – J. Phys.}, vol.~94, pp. 1--10, 2020.

\bibitem{breuer2002theory}
H.-P. Breuer and F.~Petruccione, \emph{The Theory of Open Quantum Systems}.\hskip 1em plus 0.5em minus 0.4em\relax Oxford, U.K.: Oxford Univ. Press, 2002.

\bibitem{jones2020exploring}
D.~E. Jones, B.~T. Kirby, G.~Riccardi, C.~Antonelli, and M.~Brodsky, ``Exploring classical correlations in noise to recover quantum information using local filtering,'' \emph{New J. Phys.}, vol.~22, no.~7, p. 073037, 2020.

\bibitem{barchielli2009quantum}
A.~Barchielli and M.~Gregoratti, \emph{Quantum Trajectories and Measurements in Continuous Time: The Diffusive Case}.\hskip 1em plus 0.5em minus 0.4em\relax Berlin, Germany: Springer, 2009, vol. 782.

\bibitem{netsquid}
NetSquid: The Network Simulator for Quantum Information using Discrete events [Online]. Available: \url{https://netsquid.org}.

\bibitem{qutip}
QuTiP: Quantum Toolbox in Python [Online]. Available: \url{https://qutip.org}.

\bibitem{kreuter2005experimental}
A.~Kreuter, C.~Becher, G.~P.~T. Lancaster, A.~B. Mundt, C.~Russo, H.~H\"affner \emph{et~al.}, ``Experimental and theoretical study of the 3d d\textsubscript{2}-level lifetimes of \textsuperscript{40}ca\textsuperscript{+},'' \emph{Phys. Rev. A}, vol.~71, no.~3, p. 032504, 2005.

\bibitem{wang2021single}
P.~Wang, C.-Y. Luan, M.~Qiao, M.~Um, J.~Zhang \emph{et~al.}, ``Single ion qubit with estimated coherence time exceeding one hour,'' \emph{Nat. Commun.}, vol.~12, no.~1, p. 233, 2021.

\bibitem{rancic2018coherence}
M.~Ran\v{c}i\'{c}, M.~P. Hedges, R.~L. Ahlefeldt, and M.~J. Sellars, ``Coherence time of over a second in a telecom-compatible quantum memory storage material,'' \emph{Nat. Phys.}, vol.~14, no.~1, pp. 50--54, 2018.

\bibitem{maurer2012room}
P.~C. Maurer, G.~Kucsko, C.~Latta, L.~Jiang, N.~Y. Yao \emph{et~al.}, ``Room-temperature quantum bit memory exceeding one second,'' \emph{Science}, vol. 336, no. 6086, pp. 1283--1286, 2012.

\bibitem{milul2023superconducting}
O.~Milul, B.~Guttel, U.~Goldblatt, S.~Hazanov, L.~M. Joshi \emph{et~al.}, ``Superconducting cavity qubit with tens of milliseconds single-photon coherence time,'' \emph{PRX Quantum}, vol.~4, no.~3, p. 030336, 2023.

\bibitem{reagor2016quantum}
M.~Reagor, W.~Pfaff, C.~Axline, R.~W. Heeres, N.~Ofek \emph{et~al.}, ``Quantum memory with millisecond coherence in circuit qed,'' \emph{Phys. Rev. B}, vol.~94, no.~1, p. 014506, 2016.

\bibitem{bali2024routing}
R.~Bali, A.~Tittelbaugh, S.~L. Jenkins, A.~Agrawal, J.~Horgan \emph{et~al.}, ``Routing and spectrum allocation in broadband degenerate epr-pair distribution,'' in \emph{Proc. IEEE Int. Conf. Commun. (ICC)}, 2024, pp. 4954--4960.

\bibitem{ookla2024speedtest}
{Ookla}, ``Speedtest by ookla global fixed network performance—2024-01-01 fixed tiles data,'' 2024, accessed Nov. 4, 2024. [Online]. Available: \url{https://registry.opendata.aws/speedtest-global-performance}.

\bibitem{wengerowsky2018entanglement}
S.~Wengerowsky, S.~K. Joshi, F.~Steinlechner, H.~H\"{u}bel, and R.~Ursin, ``An entanglement-based wavelength-multiplexed quantum communication network,'' \emph{Nature}, vol. 564, no. 7735, pp. 225--228, 2018.

\bibitem{sansa2022visible}
A.~Sansa~Perna, E.~Ortega, M.~Gr\"{a}fe, and F.~Steinlechner, ``Visible-wavelength polarization-entangled photon source for quantum communication and imaging,'' \emph{Appl. Phys. Lett.}, vol. 120, no.~7, p. 071102, 2022.

\end{thebibliography}

\end{document}